\documentclass[12pt]{article}
\usepackage{latexsym}
\usepackage{graphicx}
\begin{document}
\bibliographystyle{unsrt} 
 
\begin{center} 
{\LARGE \bf Exceptional Projective Geometries and Internal Symmetries}\\[10mm] 
 
Sultan Catto$^{\dag}$\\{\it Physics Department\\ The Graduate School and University Center\\ The City University of New York \\365 Fifth Avenue\\
New York, NY 10016-4309\\ and \\ Center for Theoretical Physics \\The Rockefeller University \\
1230 York Avenue\\ New York NY 10021-6399}\\[6mm] 
\end{center} 
\vbox{\vspace{5mm}}

\begin{abstract} 
A new mneumonic devise is shown to emerge in connection with $O(7)$ numerical tensors exhibiting duality and reflecting the natural $7=(4+3)$ splitting of $7$-dimensional space. Then Desargues' and Pappus' theorems are shown to be connected  through a geometry that makes use of octonionic numbers exhibiting this duality. Construction of exceptional Hilbert spaces based on Jordan algebras and exceptional projective geometries is illustrated. A brief discussion of the Moufang plane and non-Desarguesian geometries is presented. 
\end{abstract}
\vbox{\vspace{5mm}}


\vbox{\vspace{10mm}}

$^\dag$ Work supported in part by DOE contracts No. DE-AC-0276 ER 03074 and 03075, and PSC-CUNY Research Awards.
\newpage

\section*{Introduction}

There are some unique exceptional geometries which are connected with octonions and correspond to a finite number of degrees of freedom that cannot be extended. This is the new Hilbert space that we want tentatively to identify with the Hilbert space of internal symmetries carrying color and flavor quantum numbers.

One of the fundamental questions in particle physics is the understanding of the quark substructure of hadrons. We would like to understand the emergence of the hierarchy of interactions, the mass spectrum of quarks and leptons and the wider mass spread of the fundamental gauge bosons through the spontaneous breaking of a local symmetry based on a unifying group or supergroup associated with a gauge field theory.

Octonionic planes may possibly provide the geometrical foundation for the existence of internal symmetries like color and flavor. These octonionic geometries allow us to construct new finite Hilbert spaces with unique properties. The non-Desargues'ian geometric property makes them non-embeddeble in higher spaces, hence essentially finite. It also leads to peculiarities in the superposition principle in the color sector of the Hilbert space. This theory of the charge space, if correct, suggests a new geometric picture for the substructure of the material world, that of an octonionic geometry attached at each point of Einstein's Riemannian manifold for space time with local symmetry that leaves the properties of charge space invariant.

In what follows, we first introduce a simple description of octonion algebra. $O(7)$ numerical tensors will be shown to be dual and a new mneumonic devise will emerge as a useful device in their combined description. These will naturally lead into construction of two completely antisymmetric $O(8)$ tensors in $R^8$. Then, a split octonion algebra will be shown that will close into a fermionic Heisenberg algebra leading to an algebra of color forces in QCD. Application to effective hadronic supersymmetry when the split units and their conjugates become associated with quark and antiquark fields, respectively, will follow, based on $SU(m/n)$ supergroups, leading to new mass formulas describing mesons and baryons.

Last part of the paper deals with octonionic quantum mechanics and construction of exceptional Hilbert spaces based on Jordan algebras and exceptional projective geometries. Within this context a new geometric structure will be shown to emerge connecting Pappus' and Desargues' theorem with octonions and octonionic dualities. A brief discussion of the Moufang plane and its invariance group $F_4$ (that replaces $U(n) (Sp(n))$ group acting on complex (quaternionic)quantum mechanical spaces), and the non-Desarguesian geometries follow. These geometries admit as invariance groups the exceptional series ($E_6,~E_7~{\rm and} ~E_8)$. 

\vspace{2cm}

\section*{Octonion Algebra}

An octonion $x$ is a set of eight real numbers

\begin{equation}
x=(x_0,x_1,\ldots,x_7) = x_0e_0+x_1e_1+\ldots +x_7e_7
\end{equation}
that are added like vectors and multiplied according to the rules

\begin{equation}
e_0=1,~~~~e_0e_i=e_ie_0=e_i,~~~~i=0,1,\ldots,7
\end{equation}

\begin{equation}
e_\alpha e_\beta =-\delta_{\alpha\beta} + \epsilon_{\alpha\beta\gamma} e_\gamma.~~~\alpha,\beta,\gamma=1,2,\ldots,7
\end{equation}
where $e_0$ is the multiplicative unit element and $e_\alpha$'s are the imaginary octonion units. 
The structure constants $\epsilon_{\alpha\beta\gamma}$ are completely antisymmetric and take the value 1 for combinations

\begin{equation}
\epsilon_{\alpha\beta\gamma}=(165), (257), (312), (471), (543), (624), (736)
\end{equation}
Note that summation convention is used for repeated indices. 

The octonion algebra ${\cal C}$ is an algebra defined over the field ${\bf Q}$ 
of rational numbers, which as a vector space over ${\bf Q}$ has dimension 8. 
It can be described as the linear span of the units 
$e_{0},e_{1},\ldots,e_{7}$ subject to a non-associative multiplication given in its component form
in the following table:

\newpage
$$
\vbox{\offinterlineskip
\hrule
\halign{&\vrule#&
 \strut\quad\hfil#\quad\cr
height2pt&\omit&&\omit&\cr
&
$\cdot$\hfil&&$e_{1}$&&$e_{2}$&&$e_{3}$&&$e_{4}$&&$e_{5}$&&$e_{6}$&&$e_{7}
$&\cr
height2pt&\omit&&\omit&&\omit&&\omit&&\omit&&\omit&&\omit&&\omit&\cr
\noalign{\hrule}
height2pt&\omit&&\omit&&\omit&&\omit&&\omit&&\omit&&\omit&&\omit&\cr
&$e_{1}$&&-1&&$e_{3}$&&$-e_{2}$&&$e_{7}$&&$-e_{6}$&&$e_{5}$&&$-e_{4}$&\cr
\noalign{\hrule}
&$e_{2}$&&$-e_{3}$&&-1&&$e_{1}$&&$e_{6}$&&$e_{7}$&&$-e_{4}$&&$-e_{5}$&\cr
\noalign{\hrule}
&$e_{3}$&&$e_{2}$&&$-e_{1}$&&-1&&$-e_{5}$&&$e_{4}$&&$e_{7}$&&$-e_{6}$&\cr
\noalign{\hrule}
&$e_{4}$&&$-e_{7}$&&$-e_{6}$&&$e_{5}$&&-1&&$-e_{3}$&&$e_{2}$&&$e_{1}$&\cr
\noalign{\hrule}
&$e_{5}$&&$e_{6}$&&$-e_{7}$&&$-e_{4}$&&$e_{3}$&&-1&&$-e_{1}$&&$e_{2}$&\cr
\noalign{\hrule}
&$e_{6}$&&$-e_{5}$&&$e_{4}$&&$-e_{7}$&&$-e_{2}$&&$e_{1}$&&-1&&$e_{3}$&\cr
\noalign{\hrule}
&$e_{7}$&&$e_{4}$&&$e_{5}$&&$e_{6}$&&$-e_{1}$&&$-e_{2}$&&$-e_{3}$&&-1&\cr
height2pt&\omit&&\omit&&\omit&&\omit&&\omit&&\omit&&\omit&&\omit&\cr}
\hrule}
$$
\centerline{\bf Table 1: Cayley Multiplication Table}

\vskip 1.5cm

Defining the conjugates

\begin{equation}
\bar{e}_\alpha = -e_\alpha, ~~~~\bar{e}_0=e_0
\end{equation}
the octonionic conjugate $\bar{x}$ is

\begin{equation}
\bar{x}=x_0 -x_\alpha e_\alpha=x_i \bar{e}_i
\end{equation}

The scalar part of the octonion ($ Sc(x)$) and its vector part ($Vec(x)$) are

\begin{equation}
Sc(x)= \frac{1}{2} (x+\bar{x}) =x_0
\end{equation}
and

\begin{equation}
Vec(x)=\frac{1}{2}(x-\bar{x}) = x_\alpha e_\alpha
\end{equation}

Conjugate of the product of two octonions $x$ and $y$ is

\begin{equation}
(\overline{xy})=\bar{y} \bar{x}
\end{equation}
and their scalar product is defined by

\begin{equation}
<x,y>=x_i y_i=\frac{1}{2}(x\bar{y} + y \bar{x})=\frac{1}{2}(\bar{x} y + \bar{y} x)
\end{equation}
which, in terms of octonion units gives

\begin{equation}
<e_i,e_j>= \frac{1}{2} (\bar{e}_i e_j +\bar{e}_j e_i) =\frac{1}{2} (e_i \bar{e}_j + e_j \bar{e}_i)=\delta_{ij}
\end{equation}

The norm $N(x)$ of an octonion is 

\begin{equation}
N(x)=\bar{x} x = x \bar{x} = x_i x_i
\end{equation}
and is zero if $x=0$, and is always positive otherwise. For a nonzero octonion $x$, its inverse is given by 

\begin{equation}
x^{-1}= \frac{\bar{x}}{N(x)}
\end{equation}
with

\begin{equation}
(xy)^{-1} = y^{-1} x^{-1}
\end{equation}

The norm defined above satisfies

\begin{equation}
N(xy)=N(x) N(y)   \label{eq:obes}
\end{equation}

In analogous way to the quaternionic case in $R^4$,we now make the following definitions of vectorial products of octonion units, namely two new antisymmetric tensors: 

\begin{equation}
e_{ij}= \frac{1}{2} (\bar{e}_i e_j - \bar{e}_j e_i)
\end{equation}
and

\begin{equation}
e^{'}_{ij} = \frac{1}{2} (e_i \bar{e}_j - e_j \bar{e}_i)
\end{equation}
Componentwise they read as

\begin{equation}
e_{\alpha\beta} = e^{'}_{\alpha\beta} = -\epsilon_{\alpha\beta\gamma} e_\gamma
\end{equation}
and
\begin{equation}
e_{0\alpha} =-e^{'}_{0\alpha} = e_\alpha
\end{equation}
These octonionic tensors naturally enter into covariant formulation of various cross-products in $R^8$.
We can now write

\begin{equation}
\bar{e}_i e_j =\frac{1}{2} (\bar{e}_i e_j+\bar{e}_j e_i) +\frac{1}{2}(\bar{e}_i e_j - \bar{e}_j e_i)=\delta_{ij}+e_{ij}
\end{equation}

\begin{equation}
e_i \bar{e}_j =\frac{1}{2} (e_i \bar{e}_j+e_j \bar{e}_i) +\frac{1}{2}(e_i \bar{e}_j - e_j \bar{e}_i)=\delta_{ij}+e^{'}_{ij}
\end{equation}

Using the definition of commutator of two octonions

\begin{equation}
[x,y]=0
\end{equation}
the product rule implies

\begin{equation}
[e_\alpha, e_\beta]=2\epsilon_{\alpha\beta\gamma} e_\gamma
\end{equation}
and

\begin{equation}
[e_0,e_\alpha]=0
\end{equation}
so that

\begin{equation}
[x,y]= 2x_\alpha y_\beta \epsilon_{\alpha\beta\gamma} e_\gamma
\end{equation}

We see the relation between the commutator and $e_{ij}$ and $e^{'}_{ij}$ through the triality relation

\begin{equation}
[e_i, e_j] + e_{ij} + e^{'}_{ij} =0
\end{equation}

We now define the associator $ [x,y,z] $ of three octonions by

\begin{equation}
[x,y,z]= (xy) z - x (yz)
\end{equation}
which is completely antisymmetric:

\begin{equation}
[x,y,z]=[y,z,x] = [z,x,y]
\end{equation}
and

\begin{equation}
[x,y,z]=-[y,x,z]=-[x,z,y]=-[z,y,x]
\end{equation}
It is also purely vectorial since

\begin{equation}
[\overline{x,y,z}]=-[x,y,z]
\end{equation}

Since $e_0$ commutes and associates with other octonion units, only the purely vectorial parts of $ x,y,z $ contribute to the associator $ [x,y,z] $.

We now define a completely antisymmetric 4-index object $ \psi_{\alpha\beta\mu\nu} $ related to the associator as

\begin{equation}
[e_\alpha, e_\beta, e_\mu] = 2 \psi_{\alpha\beta\mu\nu} e_\nu
\end{equation}

By means of the associator and the product rules, we find

\begin{equation}
\psi_{\alpha\beta\mu\nu} =\frac{1}{2} (\delta_{\beta [ \mu} \delta_{\alpha ] \nu} + \epsilon_{\beta\gamma [\mu} 
\epsilon_{\alpha ] \gamma\nu}) 
\end{equation}

Explicit calculation of the values of $ \psi_{\alpha\beta\mu\nu} $ show it is dual to $ \epsilon_{\lambda\sigma\rho} $ 
in $R^7$. $ \psi_{\alpha\beta\mu\nu} $ take value 1 for the following combinations:

\begin{equation}
(\alpha\beta\mu\nu)=(1346), (2635), (4567), (3751), (6172), (5214), (7423)
\end{equation}

Duality property between $\epsilon_{\lambda\sigma\rho} $ and $\psi_{\alpha\beta\mu\nu}$ in $R^7$ is best seen in the following construction:

\begin{eqnarray}
\left\{
\begin{array}{ccccccl}
2 & 4 & 3 & 6 & 5 & 7 & 1  \\
5 & 7 & 1 & 2 & 4 & 3 & 6  \\
7 & 1 & 2 & 4 & 3 & 6 & 5
\end{array}
\right\}=\epsilon_{\lambda\sigma\rho}~~  \nonumber \\
\left\{
\begin{array}{ccccccl}
1 & 2 & 4 & 3 & 6 & 5 & 7 \\
3 & 6 & 5 & 7 & 1 & 2 & 4 \\
4 & 3 & 6 & 5 & 7 & 1 & 2  \\
6 & 5 & 7 & 1 & 2 & 4 & 3
\end{array}
\right\}=\psi_{\alpha\beta\mu\nu}     \label{eq:triad}
\end{eqnarray}

This table is read as

\begin{equation}
\frac{1}{2}[e_2, e_5]=e_7,~~~~\frac{1}{2}[e_4,e_7]=e_1,\ldots
\end{equation}
etc., for the triads,and

\begin{equation}
\frac{1}{2}[e_1,e_3,e_4]=-e_6,~~~~ \frac{1}{2}[e_2,e_6,e_3]=-e_5, \ldots
\end{equation}
etc., for the associators.

We see that there are seven associative planes and their transverse non-associative planes. While an associative plane is closed under commutation, a non-associative plane closes under the associator.

We now present a new diagrammatic representations of octonionic multiplication tables on two consecutively drawn circles below. Diagrams below show the triad $\epsilon_{165}$ and its associated 4-index object $\psi_{7423}$, corresponding to the last line in Eq.(\ref{eq:triad}). While six successive rotation of the triangle on the left circle through angle $\frac{2\pi}{7}$ shows the multiplication rule for the associated triads, the arrow-like figure on the right circle produces its dual part also through the same successive rotations. We receive the relations and multiplication rules shown earlier by simultaneous successive clockwise (or counterclockwise) rotations of the dials, while keeping the triangle on the left figure and the arrow-like figure on the right one intact. Here we also note that, along with the identity $e_0=1$, the elements corresponding to the corners of the triangle in the left figure form a basis of of a $SU(2)$ quaternion algebra.

\vskip 2.0cm
\begin{center}
\unitlength=.3mm
\begin{picture}(240,80)(0,30)
\put(20,70){\circle{150}}
\put(20,95){\circle*{3.7}}
\put(40,85){\circle*{3.7}}
\put(0,85){\circle*{3.7}}
\put(44,65){\circle*{3.7}}
\put(-4,65){\circle*{3.7}}
\put(35,50){\circle*{3.7}}
\put(5,50){\circle*{3.7}}
\end{picture}
\end{center}
\begin{center}
\unitlength=.3mm
\begin{picture}(20,-120)(0,0)
\put(20,70){\circle{150}}
\put(20,95){\circle*{3.7}}
\put(40,85){\circle*{3.7}}
\put(0,85){\circle*{3.7}}
\put(44,65){\circle*{3.7}}
\put(-4,65){\circle*{3.7}}
\put(35,50){\circle*{3.7}}
\put(5,50){\circle*{3.7}}
\end{picture}
\end{center}
\begin{center}
\unitlength=.3mm
\begin{picture}(240,180)(0,0)
\put(0,283){\vector(4,0){35}}
\put(-2.1,262){\vector(1,4){5}}
\put(40,283){\vector (-2,-1){40}}
\put(145,248){\vector(1,2){7}}
\put(130,289){\vector(1,-3){14}}
\put(150,260){\vector(-3,-1){33}}
\put(115,247){\vector(1,0){29}}
\put(20,305){\makebox(0,0){$e_7$}}
\put(130,305){\makebox(0,0){$e_7$}}
\put(-10,285){\makebox(0,0){$e_5$}}
\put(100,285){\makebox(0,0){$e_5$}}
\put(50,285){\makebox(0,0){$e_1$}}
\put(160,285){\makebox(0,0){$e_1$}}
\put(-14,265){\makebox(0,0){$e_6$}}
\put(96,265){\makebox(0,0){$e_6$}}
\put(54,265){\makebox(0,0){$e_2$}}
\put(164,265){\makebox(0,0){$e_2$}}
\put(-2,240){\makebox(0,0){$e_3$}}
\put(110,240){\makebox(0,0){$e_3$}}
\put(42,240){\makebox(0,0){$e_4$}}
\put(154,240){\makebox(0,0){$e_4$}}

\end{picture}
\end{center}
\centerline{\bf Fig. 1}
\vspace{2cm}

We now put the two pictures together:
\begin{center}
\includegraphics[width=80mm]{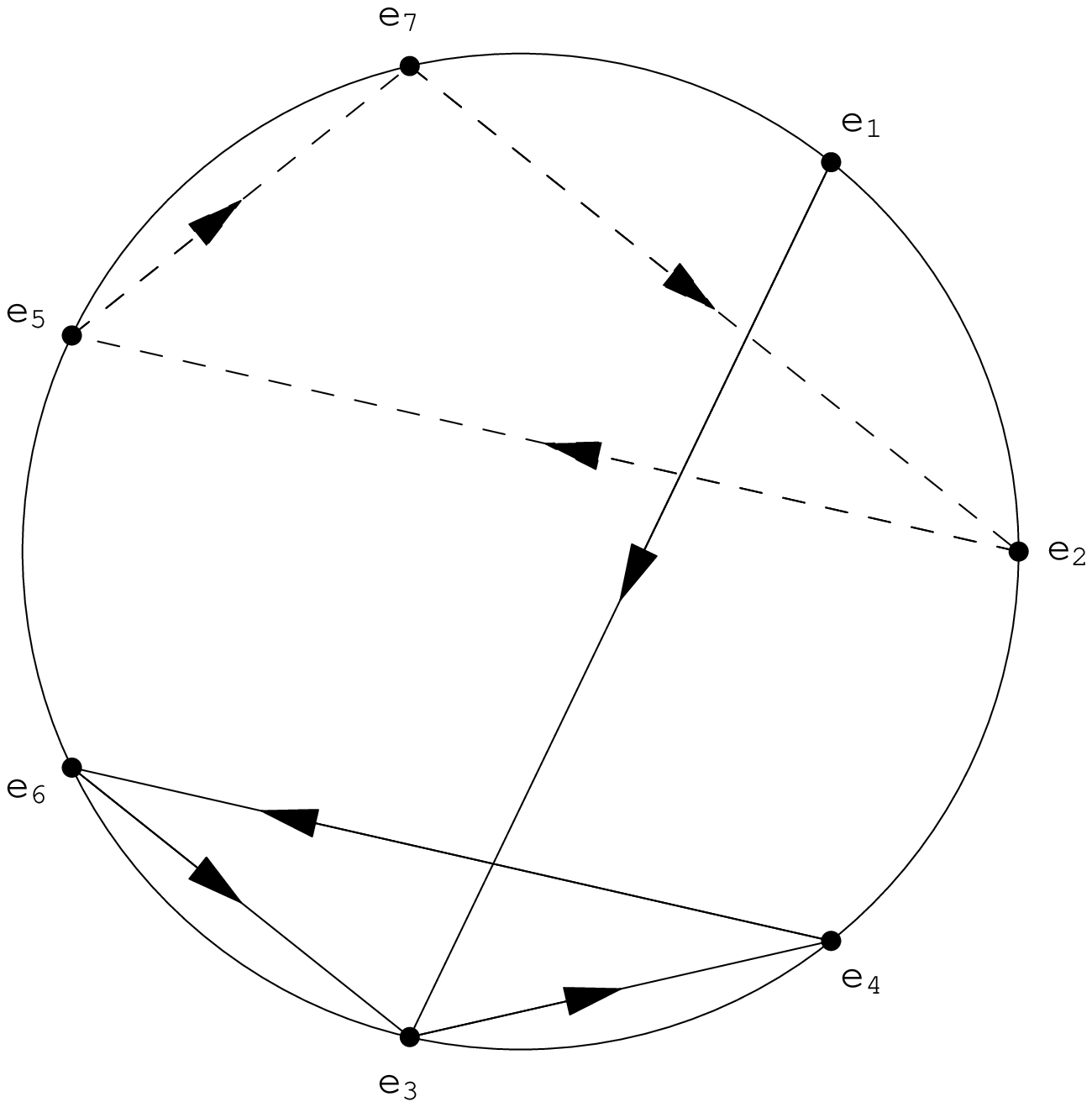}
\end{center}
\centerline{\bf Fig. 2}

\vskip 1.5cm

Keeping the solid-lined distorted arrow like figure and the triangle formed by the dashed lines intact, the successive rotations of the dial at increments $2\pi/7$ accounts for all the symmetries between $\epsilon_{ijk}$'s and the $\psi_{\alpha\beta\gamma\delta}$'s given earlier.

We note that the associator plays an important role in the derivation of various properties of octonions. Lets take a look for example at the composition property of the norm for real octonions. Since $[e_\alpha,e_\beta,e_\gamma]=2\psi_{\alpha\beta\gamma\delta}~e_\delta$ is purely vectorial, we deduce that Sc$(e_\alpha(e_\beta e_\gamma))$=Sc$((e_\alpha e_\beta)e_\gamma)$ or Sc$(a(bc))$=Sc$((ab)c)$, so that

\begin{eqnarray}
N(ab)&=&Sc[(ba)(ab)]=Sc[b(a(ab))] \nonumber \\
 &= & Sc[(baa)b]=N(a)~Sc(bb)=N(a)N(b)
\end{eqnarray}
which is same as that of Eq.(\ref{eq:obes}).

By way of writing down the three fundamental quadrilinear Moufang identities$^{\cite{gurtze},\cite{yokota}}$ it can easily be shown that 

\begin{equation}
[e_\alpha,e_\beta,e_\gamma]=2\psi_{\alpha \beta \gamma \delta}~ e_\delta = 2 \epsilon_{[\alpha\beta}^\kappa \epsilon_{\kappa ]\gamma}^\kappa~e_\gamma  
\end{equation}
which defines a fully antisymmetric $4$ index object $\psi_{\alpha\beta\gamma\delta}$. Reflecting the natural $7=(4+3)$ splitting of $7$-dimensional space, the $\psi_{\alpha\beta\gamma\delta}$ is dual to the $\epsilon_{\alpha\beta\gamma}$. From $O(7)$ numerical tensors $\psi_{\alpha\beta\gamma\delta}$ and  $\epsilon_{\alpha\beta\gamma}$, one can now construct two completely antisymmetric $O(8)$ tensors in $R^8$. A detailed description will be given elsewhere$^{\cite{catmor}}$.

\section*{Split Octonion Algebra} 

The Cayley algebra over the field of complex numbers

\begin{equation}
{\cal C}_{\bf C}={\cal C} \otimes_{\bf Q}{\bf C}
\end{equation}
is a split algebra, that is to say if has zero divisors. As a basis for 
${\cal C}_{\bf C}$ we can take

\begin{equation}
u_{1}={1 \over 2}(e_{1}+ie_{4}),\quad u_{1}^{*}={1 \over 2}(e_{1}-ie_{4})
\end{equation}

\begin{equation}
u_{2}={1 \over 2}(e_{2}+ie_{5}),\quad u_{2}^{*}={1 \over 2}(e_{2}-ie_{5})
\end{equation}

\begin{equation}
u_{3}={1 \over 2}(e_{3}+ie_{6}), \quad u_{3}^{*}={1 \over 2}(e_{3}-ie_{6})
\end{equation}

\begin{equation}
u_{0}={1 \over 2}(1+ie_{7}),\quad u_{0}^{*}={1 \over 2}(1-ie_{7})
\end{equation}
The multiplication table for the split complex Cayley algebra

\begin{equation}
{\cal C}_{\bf C}={\bf C}u_{1} \oplus {\bf C}u_{2} \oplus {\bf C}u_{3} \oplus 
{\bf C}u_{0} 
\oplus {\bf C}u_{1}^{*} \oplus {\bf C}u_{2}^{*} \oplus {\bf C}
{\bf C}
u_{3}^{*} \oplus {\bf C}
u_{0}^{*}  
\end{equation}
It can be verified easily that the multiplication table for the split Cayley 
algebra is
\vskip 1cm

$$
\vbox{\offinterlineskip
\hrule
\halign{&\vrule#&
 \strut\quad\hfil#\quad\cr
height2pt&\omit&&\omit&\cr
& $\cdot$\hfil&&$u_{0}^{*}$&&$u_{1}^{*}$&&$u_{2}^{*}$&&$u_{3}^{*}$&&$u_{0}$
                                        &&$u_{1}$&&$u_{2}$&&$u_{3}$&\cr
height2pt&\omit&&\omit&&\omit&&\omit&&\omit&&\omit&&\omit&&\omit&&\omit&\cr
\noalign{\hrule}
height2pt&\omit&&\omit&&\omit&&\omit&&\omit&&\omit&&\omit&&\omit&&\omit&\cr
&$u_{0}^{*}$&&$u_{0}^{*}$&&$u_{1}^{*}$&&$u_{2}^{*}$&&$u_{3}^{*}$&&0&&0&
&0&&0&\cr
\noalign{\hrule}
&$u_{1}^{*}$&&0&&0&&$u_{3}$&&$u_{2}^{*}$&&$u_{1}^{*}$&&$-u_{0}^{*}$&&0&&0
&\cr
\noalign{\hrule}
&$u_{2}^{*}$&&0&&$-u_{3}$&&0&&$u_{1}$&&$u_{2}^{*}$&&0&&$-u_{0}^{*}$&&
0&\cr
\noalign{\hrule}
&$u_{3}^{*}$&&0&&$-u_{2}$&&$-u_{1}$&&0&&$u_{3}^{*}$&&0&&0&&$-u_{0}^{*}
$&\cr
\noalign{\hrule}
&$u_{0}$&&0&&0&&0&&0&&$u_{0}^{*}$&&$u_{1}^{*}$&&$u_{2}$&&$u_{3}$&\cr
\noalign{\hrule}
&$u_{1}$&&$u_{1}$&&$-u_{0}$&&0&&0&&0&&0&&$u_{3}^{*}$&&$-u_{2}^{*}$&\cr
\noalign{\hrule}
&$u_{2}$&&$u_{2}$&&0&&$-u_{0}$&&0&&0&&$-u_{3}^{*}$&&0&&$u_{1}^{*}$&\cr
\noalign{\hrule}
&$u_{3}$&&$u_{3}$&&0&&0&&$-u_{0}$&&0&&$u_{2}^{*}$&&$-u_{1}^{*}$&&0&\cr
height2pt&\omit&&\omit&&\omit&&\omit&&\omit&&\omit&&\omit&&\omit&\cr}
\hrule}
$$

\centerline{\bf Table 2: Split Cayley Multiplication Table}

\vspace{1cm}
The split Cayley algebra over ${\bf C}$ actually has a ${\bf Q}$-structure 
which for ease of notation we will simply denote by ${\cal C}$ and has the 
property that 

\begin{equation}
{\cal C}\otimes_{\bf Q}{\bf C}= \hbox{Split Cayley Algebra over} {\bf ~C}
\end{equation}

As this rational ${\bf Q}$-structure is the main object of interest, we now 
proceed to describe it in terms of $2 \times 2$ matrices.
\vskip .5cm

Let $F$ be a field of characteristic $\neq 2$ and let $V={\bf Q}^{3}$ be 
Euclidean three-space over the field of rational numbers. The split Cayley 
algebra over $F$ (given in terms of Zorn's matrices) is defined by 

\begin{equation}
{\cal C}= \{ \left( \matrix{\alpha & a \cr
                                b & \beta \cr} \right) : \alpha, \beta \in 
{\bf Q}, \hbox{ and } a,b \in V_{\bf Q} \}
\end{equation}
The bilinear multiplication of two Zorn matrices is defined by 

\begin{equation}
\left(\matrix{\alpha&a \cr
                b &\beta \cr}\right) 
  \left(\matrix{\gamma&c \cr
                d &\delta \cr} \right)=
 \left(\matrix{\alpha \gamma+a \cdot d & \alpha c+\delta a- b \times d \cr
               \gamma b+\beta d+a \times c & \beta \delta +b \cdot c \cr} 
\right)
\end{equation}
The matrix 

\begin{equation}
I=\left(\matrix{1 &0 \cr
                        0 &1 \cr} \right)
\end{equation}
clearly acts as an identity element for this algebra. To obtain the split Cayley algebra over an arbitrary field $F$, we need 
only replace the field ${\bf Q}$ in the above construction by $F$, i.e. $V={\bf Q}^{3}$ becomes $V=F^{3}$, etc.
\vskip .5cm

An explicit isomorphism between ${\cal C} \otimes_{\bf Q}{\bf C}$ and 
${\cal C}_{\hbox{split}}$ is given by the correspondence

\begin{equation}
 {\cal C}_{\hbox{split}} \rightarrow {\cal C} \otimes_{\bf Q}{\bf C}
\end{equation}

\begin{equation}
u_{0} \mapsto \left(\matrix{0&0 \cr
                                0&1 \cr}\right),\quad
                u_{0}^{*} \mapsto \left(\matrix{1&0 \cr
                                        0&0 \cr}\right)
\end{equation}

\begin{equation}
u_{i} \mapsto \left(\matrix{0&0 \cr
                e_{i} &0 \cr}\right),\quad u_{i}^{*} \mapsto 
\left(\matrix{0& -e_{i} \cr
                        0&0 \cr}\right),\quad i=1,2,3
\end{equation}

Under this mapping the conjugation of Cayley numbers carries over into a 
natural involution on the set of Zorn matrices:

\begin{equation}
A=\left(\matrix{\alpha & a\cr
                b & \beta \cr}\right) \mapsto \overline{A}=
\left( \matrix{\beta & -a \cr
                        -b & \alpha \cr}\right)
\end{equation}
One verifies easily that 

\begin{equation}
A \overline{A}=\overline{A} A=(\alpha \beta-a \cdot b)I
\end{equation}
The expression

\begin{equation}
q(A)=\alpha \beta -a \cdot b
\end{equation}
defines a quadratic form admitting composition, i.e. 

\begin{equation}
q(A \cdot B)=q(A) \cdot q(B)
\end{equation}
for all $A,B \in {\cal C}$.
The symmetric bilinear form obtained by polarization

\begin{equation}
2(A,B)=q(A+B)-q(A)-q(B)
\end{equation}
is non-degenerate. It is also inmediate that ${\cal C}$ has zero divisors, 
e.g.

\begin{equation}
\left(\matrix{0&a \cr 
                0&0 \cr}\right)^{2}=0,\quad a \in V
\end{equation}
The general theory of algebras ascertains that there is only one non-
associative algebra with unit over a field $F$ carrying a non-degenerate 
quadratic form admitting composition and possessing zero-divisors. The 
description above of the split Cayley algebra will be useful to us when 
determining the derivation algebra of ${\cal C}$. For further details we refer the reader to our paper "Exceptional Quantum Mechanical Spaces and Their Invariance Groups: Physical Interpretation," S. Catto and Carlos Moreno, to be published$^{\cite{catmor}}$.

To compactify our notation, we write

\begin{equation}                      
u_{0} = \frac{1}{2} (1 +i e_{7}) ,                      
~~~~~~~u_{0}^{*} = \frac{1}{2} (1 -i e_{7})  \label{eq:ellisekiz}                     
\end{equation}                        

\begin{equation}                      
u_{j} = \frac{1}{2} (e_{j} +i e_{j+3}) ,                      
~~~~~u_{j}^{*} = \frac{1}{2} (e_{j} -i e_{j+3}) , ~~~j=1,2,3  \label{eq:ellidokuz}                    
\end{equation}                       
where $i=\sqrt{-1}$ commutes with any $e_\alpha$.
                      
The automorphism group of the octonion algebra is the 14-parameter                       
exceptional group $G_{2}$.  The imaginary octonion units                       
$e_{\alpha} (\alpha  =1,...,7)$ fall into its 7-dimensional representation.                      
                       
   Under the $SU(3)^{c}$ subgroup of $G_{2}$ that leaves $e_{7}$                       
invariant, $u_{0}$ and $u_{0}^{*}$ transform like singlets, while $u_{j}$ and                       
$u_{j}^{*}$ transform like a triplet and antitriplet respectively.  The multiplication table can now be                       
written in a manifestly $SU(3)^{c}$ invariant manner (together with the                       
complex conjugate equations):

\begin{equation}                      
u_{0}^{2} = u_{0},~~~~~u_{0}u_{0}^{*} = 0                      
\end{equation}                       

\begin{equation}                      
u_{0} u_{j} = u_{j} u_{0}^{*} = u_{j},~~~~~                      
u_{0}^{*} u_{j} = u_{j} u_{0} = 0                          
\end{equation}                       

\begin{equation}                      
u_{i} u_{j}  = - u_{j} u_{i} = \epsilon_{ijk} u_{k}^{*}                            
\end{equation}                       

\begin{equation}                      
u_{i} u_{j}^{*} =  - \delta_{ij} u_{0}                               
\end{equation}                       
where $\epsilon_{ijk}$ is completely antisymmetric with  $\epsilon_{ijk} =1$                      
for  $ijk$  = $123$, $246$, $435$, $651$, $572$, $714$, $367$.                        

From these we deduce                       

\begin{equation}                      
(u_{i} u_{j}) u_{k} = - \epsilon_{ijk} u_{0}^{*}                       
\end{equation}

Now, we put together the compactified multiplication table for the split octonion units:
\vskip 1.0cm

\begin{center}
\begin{tabular}{|l|c|c|c|c|}\hline
 & $u_0$ & $u_{0}^{*}$ & $u_{k} $ & $u_{k}^{*}$  \\ \hline 
$u_0$  &  $ u_0 $  & $0$ & $u_k$ & $0$ \\ \hline

$u_{0}^{*}$  &  $ 0 $  & $u_{0}^{*}$ & $0$ & $u_{k}^{*}$ \\ \hline

$u_j$  &  $ 0 $  & $u_{j}$ & $\epsilon_{jki}u_{i}^{*}$ & $-\delta_{jk}u_{0}$ \\ \hline

$u_{j}^{*}$  &  $ u_{j}^{*} $  & $0$ & $-\delta_{jk}u_{0}^{*}$ & $\epsilon_{jki}u_{i}$ \\ \hline

\end{tabular}
\end{center}
\centerline{\bf Table 3}

\vskip 1.0cm
It is worth noting that $u_i$ and $u_{j}^{*}$ behave like fermionic annihilation and creation operators:

\begin{equation}
\{u_i,u_j\}=\{u_{i}^{*},u_{j}^{*}\}=0,~~~\{u_i,u_{k}^{*}\}=-\delta_{ij}
\end{equation}                      

This fermionic Heisenberg algebra shows the three split units $u_i$ to be Grassmann numbers. Being non-associative, these split units give rise to an exceptional Grassmann algebra.

Operators $u_i$, unlike ordinary fermion operators, are not associative. We also have 
$\frac{1}{2}[u_i,u_j]=\epsilon_{ijk}~u_k^{*}$. The Jacobi identity
does not hold since 

\begin{equation}
[u_i,[u_j,u_k]]=- i e_7 \neq 0
\end{equation}
where $e_7$, anticommute with $u_i$ and $u_i^{*}$.

We note that, like the imaginary units $e_\alpha$, the split units cannot be represented by matrices. Unlike the octonion algebra, the split octonion algebra contains zero divisors and is therefore not a division algebra.

The associators of split octonion units are given below:

\begin{equation}
[u_i,u_j,u_k]=\epsilon_{ijk} (u_{0}^* -u_0)
\end{equation}
 
\begin{equation}
[u_i^*,u_j^*,u_k^*]=\epsilon_{ijk} (u_0-u_{0}^*)
\end{equation}
 
\begin{equation}
[u_i,u_j,u_0]=-\epsilon_{ijk} u_k^* 
\end{equation}
 
\begin{equation}
[u_i,u_j,u_0^*]=\epsilon_{ijk} u_k^* 
\end{equation}
 
\begin{equation}
[u_i,u_j,u_k^*]=\delta_{jk} u_i - \delta_{ik} u_j 
\end{equation}
 
\begin{equation}
[u_i,u_j^*,u_k^*]=\delta_{ki} u_j^* - \delta_{ij} u_k^* 
\end{equation}
 
\begin{equation}
[u_i^*,u_j^*,u_0]=\epsilon_{ijk} u_k 
\end{equation}
 
\begin{equation}
[u_i^*,u_j^*,u_0^*]=-\epsilon_{ijk} u_k 
\end{equation}
 
\begin{equation}
[u_i,u_j^*,u_0]=0 
\end{equation}

\begin{equation}
[u_i,u_j^*,u_0^*]=0 
\end{equation}
 
Defining hermitian conjugation as both complex and octonionic conjugation we write
 
\begin{equation}
u_{i}^{\dagger} = {\bar{u}}_i^* =- u_i^*,~~~~~ u_{0}^{\dagger} = {\bar{u}}_0^* = u_0
\end{equation}

We also make new definitions:
 
\begin{equation}
u_{\mu\nu} = \frac{1}{2} (u_{\mu}^{\dagger} u_{\nu} - u_{\nu}^{\dagger} u_{\mu})
\end{equation}
and

\begin{equation}
u^{'}_{\mu\nu} = \frac{1}{2} (u_{\mu} u_{\nu}^{\dagger} - u_{\nu} u_{\mu}^{\dagger})
\end{equation}
and see that the left handed product

\begin{equation}
u'_{\mu\nu} =0
\end{equation}
and, in the component form the right handed product $u'_{\mu\nu}$ survives
only as $u_{0i}=\frac{1}{2} e_i$, i.e.
 
\begin{equation}
u_{ij}=0
\end{equation}
with

\begin{equation}
u_{0i}=\frac{1}{2} (u_i +u_i^*) = \frac{1}{2} e_i
\end{equation}
thereby reducing the octonions to purely vectorial quaternions. Next, we give some brief applications of the octonions in effective supersymmetric field theories and in quantum mechanical description of Hilbert space of internal symmetries. A new connections are found on connecting Pappus' and Desargues' geometries through octonionic projective geometries and their applications to dualities will be shown below.

\section*{Effective Dynamical Supersymmetry: A brief discussion}

A phenomenological manifestation of octonionic structure (based on Cayley numbers) has been found in connection with quark dynamics inside hadrons leading to an effective dynamical supersymmetry$^{\cite{1}, \cite{2}}$. It is based on $SU(m/n)$ supergroups yielding combined classification of mesons and baryons. Under the color group $SU(3)^c$, $q\bar{q}$ and $qq$ states transform as

\begin{equation}
q\bar{q}:~~~ \mbox{\boldmath$3$} \times \bar{\mbox{\boldmath$3$}}=\mbox{\boldmath$1$}+\mbox{\boldmath$8$}~;~~~~~~~qq:~~~
\mbox{\boldmath$3$} \times \mbox{\boldmath$3$} =\bar{\mbox{\boldmath$3$}} +\mbox{\boldmath$6$}   \label{eq:one}  
\end{equation}

Under the spin-flavor group $SU_{sf}(6)$, they transform as $~qq:~~\mbox{\boldmath$6$}\times \bar{\mbox{\boldmath$6$}}=\mbox{\boldmath$1$}+\mbox{\boldmath$35$}$, and $~qq:~~\mbox{\boldmath$6$}\times \mbox{\boldmath$6$}=\mbox{\boldmath$15$}+\mbox{\boldmath$21$} $, respectively. Now, mesons are $q\bar{q}$ states while baryons are $qqq$ states. If one writes $qqq$ as $qD$, where $D$ is a diquark, $D=qq$, the quantum numbers of $D$ are: for color, $\bar{\mbox{\boldmath$3$}}$, since when combined with $q$ must give a color singlet; for spin flavor, $\mbox{\boldmath$21$}$, since when combined with color must give antisymmetric wave functions. But the quantum numbers of $\bar{q}$ for color is $\bar{\mbox{\boldmath$3$}}$, and for spin-flavor, $\bar{\mbox{\boldmath$6$}}$. Thus $\bar{q}$ and $D$ have the same color quantum numbers (color forces can not distinguish between $\bar{q}$ and $D$). Therefore there is an approximate dynamic supersymmetry in hadrons with supersymmetric partners

\begin{eqnarray}
\psi= \left( \begin{array}{c} \bar{q} \\ D   \end{array} \right)~,~~~~~~\bar{\psi}=\left( \begin{array}{c} q \\ \bar{D}   \end{array} \right)    
\end{eqnarray}
            
All hadrons can be obtained by combining $\psi$ and ${\bar{\psi}}^T$: mesons are $q\bar{q}$, baryons are $qD$, antibaryons are $\bar{q}\bar{D}$ and exotic mesons are $D\bar{D}$. Since the dimensions of the internal degrees of freedom of $s=1/2$ quarks is $6$, and of diquarks for $s=1$, 18 and for $s=0$ is $3$, then the corresponding supersymmetry is $SU_{sf}(6/21)$.

In a way similar to that used in the construction of mass formulas for $SU_f(3)$ and $SU_{sf}(6)$ one can also construct mass formulas for $SU_{sf}(6/21)$ that describe simultaneously mesons and baryons. Important consequences of these formulas are: (i) slopes of mesonic and baryonic trajectories are identical; (ii) masses of some particles satisfy particular relations, for example $m_\Delta^2-m_N^2=\frac{9}{8}(m_\rho^2-m_\pi^2)$, described by semirelativistic and relativistic formulations. Both of these are experimentally verified. $SU_{sf}(6/21)$ supersymmetry also naturally leads to the existence of $D\bar{D}$ states. There are now several indications that $a_0(980)$ and $f_0(975)$ mesons are quark-antidiquark states. Review of this model and recent experimental situation is discussed in detail in G\"ursey and Tze$^{\cite{gurtze}}$, Anselmino, et al$^{\cite{3}}$, and Klempt$^{\cite{4}}$.

It is worth noting that $u_i$ and $u_{j}^{*}$ behave like fermionic annihilation and creation operators:
$ \{u_i,u_j\}=\{u_{i}^{*}$, $u_{j}^{*}\}=0$, and $\{u_i,u_{k}^{*}\}=-\delta_{ij}$. This fermionic Heisenberg algebra shows the three split units $u_j$ to be Grassmann numbers. Being non-associative, these split units give rise to an exceptional Grassmann algebra. Thus this algebra allows: triplet$\times$triplet=antitriplet, and triplet$\times$antitriplet=singlet. Now the multiplication rules for color is given in Eq.(\ref{eq:one}). But $\mbox{\boldmath$8$}$ is not allowed if we want color singlet mesons, and $\mbox{\boldmath$6$}$ is not allowed if we want to have color singlet baryons. Thus the multiplication rules of quarks and antiquarks are identical to those of split octonions. Therefore associating quark fields with split octonion units $u_i$ by choosing $q_\alpha=u_i q_\alpha^{~i}=\mbox{\boldmath$u$}\cdot {\mbox{\boldmath$q$}}_\alpha$, and antiquarks by $\bar{q}_\beta=u_i^\dag \bar{q}_\beta^{~j}=-{\mbox{\boldmath$u$}}^* \cdot {\bar{\mbox{\boldmath$q$}}}_\beta$, an effective supersymmetric model is constructed based on octonions where all the unwanted states are automatically  suppressed by means of this algebra. This model gave the first application in particle physics of dynamic (internal) supersymmetries. For a brief history of dynamic symmetries we refer to an article by Franco Iachello$^{\cite{6}}$, and for an extensive review of octonions to John Baez's comprehensive article$^{\cite{7}}$.

\section*{Octonions and Exceptional Hilbert Spaces}

~~~~~~~~In recent years a careful study of the symmetries of the directly 
observed hadronic world and strong interactions led us to the colored quark 
substructure$^{\cite{fgmg}, \cite{sc}}$. Also, the properties of weak and 
electromagnetic interactions help us to unveil a deep symmetry between 
the directly observed leptons and the partly observable quarks. These 
fermions seem to be represented by local fields that belong to multiplets 
of a single unifying group $G$ that is valid locally. What distinguishes this group $G$?  What is so basic about the exact color group? In trying to answer these questions we move to the next sphere of abstraction and see if the unification of strong, electromagnetic and weak interactions points to the geometrical properties of a deeper Hilbert space structure.
         
         During the ebullient first decade after the birth of Quantum 
Mechanics, the grand masters of the period explored all the possibilities 
for the extension of Quantum Mechanical concepts and methods. In 1933 
Jordan$^{\cite{jor}}$  suggested a new formulation of Quantum Mechanics based on 
what became to be known as Jordan algebras. In the standard formulation, 
states are represented by vectors (kets) in a Hilbert space, while observables 
are represented by Hermitian matrices acting on the states. In the Jordan 
formulation an algebra of observables is defined so that observables can be 
combined to give other observables. For instance, in classical mechanics if 
$p$ and $q$ are observables, so are $p^{2}$ , $q^{2}$ and $pq$. If $p$ and 
$q$ are replaced by hermitian matrices $P$ and $Q$, the only the combination

\begin{equation}
         P\cdot Q = Q\cdot P =\frac{1}{2}(P Q + Q P)     \label{eq:bir}                   
\end{equation}
is a hermitian matrix, hence an observable representing the classical $pq$. 
The symmetrized product Eq.(\ref{eq:bir}) is the Jordan product of $P$ and $Q$. 
Unlike ordinary matrix multiplication which is noncommutative but associative, the 
Jordan product is commutative but not associative. The associator

\begin{equation}
         [A,B,C] = (A\cdot B)\cdot C - A\cdot (B\cdot C)   
\end{equation}
is antisymmetrical in $A$ and $C$ and satisfies the Jordan identity

\begin{equation}         
         [A,B,A^{2}]=0,                         
\end{equation}
which makes the algebra of observables power associative so that
$P^{n}$ is an unambiguous expression that represents the classical observable 
$p^{n}$.
         
	In the Jordan scheme the states are also represented by certain 
hermitian matrices $P_{\alpha}$ associated with the kets $\mid \alpha>$.  They 
are the projection operators

\begin{equation}         
        P_\alpha=\mid \alpha><\alpha \mid                                
\end{equation}
that project the state $\mid \alpha>$. They satisfy the relations

\begin{equation}       
        P^{2}_\alpha=P_\alpha, \hspace{15pt} Tr P_\alpha=1   
\end{equation}
         
         Now, all the measurable quantities in quantum mechanics are transition probabilities of the form

\begin{equation}       
         \Pi_{\alpha\beta}(\Omega)=|<\alpha \mid \Omega|\beta> \mid^{2},               
\end{equation}
where $\Omega$ is a hermitian operator and

\begin{equation}       
         \Omega_{\alpha\beta}=<\alpha \mid \Omega \mid \beta>                       
\end{equation}
is the matrix element of the observable $\Omega$ between the states $\mid \alpha>$
and $\mid \beta>$. We can write$^{\cite{jp}}$

\begin{equation}       
\Pi_{\alpha\beta}(\Omega)=<\alpha|\Omega|\beta><\beta|\Omega|\alpha>=Tr 
(\Omega P_\beta\Omega P_\alpha)   
\end{equation}
We have to show that the expression between brackets can be expressed purely 
by means of the Jordan product. Let

\begin{equation}      
         U(\Omega)P_\beta=\Omega P_\beta\Omega                                
\end{equation}
We have

\begin{equation}        
         \Pi_{\alpha\beta}(\Omega)=Tr\{P_\alpha\cdot U(\Omega)P_\beta\},       
\end{equation}
where the dot refers to the Jordan product.
        
	Furthermore we have the identity

\begin{equation}        
         \{ABC\}=\frac{1}{2}(ABC+CBA)=(A\cdot B)\cdot C+A\cdot (B\cdot C)-(A\cdot C)\cdot B
\end{equation}
so that as a special case

\begin{equation}        
         \Omega P_\beta\Omega=\{\Omega P_\beta\Omega\}=2(\Omega\cdot P_\beta)\cdot 
\Omega-\Omega^{2}\cdot P_\beta                                      
\end{equation}
It follows that

\begin{equation}        
         U(\Omega)P_\beta=\{\Omega P_\beta\Omega\}                           
\end{equation}
where the right-hand side is expressed by the Jordan product only. Hence the 
absolute square of the matrix element $\Omega_{\alpha \beta}$ is

\begin{equation}                 
\Pi_{\alpha\beta}(\Omega)={\rm Tr}(P_\alpha\cdot U(\Omega)P_\beta)={\rm Tr}(P_\alpha\cdot
\{\Omega P_\beta\Omega\})={\rm Tr}(U(\Omega)P_\alpha\cdot P_\beta) \label{eq:yedi}   
\end{equation}
         
         In 1934, Jordan, von Neumann and Wigner$^{\cite{jnw}}$ showed that the 
Jordan algebras are always in one-to-one correspondence with the ordinary 
matrix algebra over complex numbers with one exception. There is an 
exceptional Jordan algebra $J^{8}_{3}$ of $3 \times 3$ matrices over octonions 
which are hermitian with respect to octonion conjugation.  They have the form

\begin{equation}
      \begin{array}{ccc} 
	J &
	= &
	\left(\begin{array}{ccc}
	\alpha & c & \bar{b} \\
	\bar{c} & \beta & a \\
	b & \bar{a} & \gamma
	\end{array}\right)
	\end{array}              \label{eq:uc}
\end{equation}
Where the bar denotes octonion conjugation $\alpha,\beta,\gamma$ are 
self-conjugate. Then Albert$^{\cite{alb}}$ proved $J^{8}_3$ was the only exceptional 
Jordan algebra.
	The octonion algebra is neither commutative nor associative, but 
because of $N (ab)= N(a)N(b)$ it is a normed composition algebra which together with 
real numbers, complex numbers and quaternions form the only normed composition 
algebras (the four Hurwitz division algebras). The algebra is alternative, 
as the associator

\begin{equation}       
         [abc]=(ab)c-a(bc)                    
\end{equation}
is antisymmetric in $a$, $b$ and $c$. The scalar part of the octonion $a$ is defined as

\begin{equation}
       {\rm Sc}~a = \frac{1}{2}(a+\bar{a})=a_o.               
\end{equation}
The diagonal elements $\alpha,\beta,\gamma$ of $J$ in (\ref{eq:uc}) are scalar
	 in this sense.
         
 A complex octonion is still of the form $\bar{a}=a_0-e_\alpha a_\alpha$ but with complex 
components. The resulting complex $J$ is still an element of an exceptional 
Jordan algebra although the complex octonions do not form a division algebra.
         
         To construct the equivalent of projection operators we introduce 
the Freudenthal product$^{\cite{freud}}$

\begin{eqnarray}         
 J_1\times J_2&=& J_2\times J_1=J_1\cdot J_2-\frac{1}{2}J_1 {\rm Tr} J_2 -\frac{1}{2}
J_2 {\rm Tr} J_1  \nonumber \\
&-& \frac{1}{2}I({\rm Tr} J_1\cdot J_2-{\rm Tr} J_1~~ {\rm Tr} J_2) 
\end{eqnarray}
The expression

\begin{equation}        
         {\rm Det}(J)=\frac{1}{3}{\rm Tr}(J\cdot(J \times J))                       
\end{equation}
is the determinant of $J$.  Matrices $P \in J^{8}_3$ which obey the conditions

\begin{equation}
        P \times P = O                              
\end{equation}
and

\begin{equation}         
         {\rm Tr} P=1               \label{eq:bes}
\end{equation}
can be shown to satisfy also

\begin{equation}       
              P^{2}=P                       
\end{equation}
so that they are projection operators. They have the form

\begin{equation}
P_\alpha 
	= 		\left( \begin{array}{c}
 	\alpha_1 \\
	\alpha_2 \\
	\alpha_3 
	\end{array}\right)
\begin{array}{ccc} 
(\bar\alpha_1 & \bar\alpha_2 & \bar\alpha_3) \\
    &    &    \\
     &    &    
	\end{array}            \label{eq:alti}
\end{equation}
where $\alpha_1, \alpha_2, \alpha_3$ are three octonions one of which is 
self-conjugate. Eq.(\ref{eq:bes}) is satisfied if

\begin{equation}       
 \alpha_1\bar\alpha_1+\alpha_2\bar\alpha_2+\alpha_3\bar\alpha_3 =1      
\end{equation}
         
         $P_\alpha$ defined by Eq.(\ref{eq:alti}) is a generalization of the 
projection operator Eq.(\ref{eq:yedi}) in the associative case. Hence matrix 
elements of an operator $J$ can be defined analogously to Eq.(\ref{eq:yedi}) by

\begin{equation}        
         \Pi_{\alpha\beta}(J)=Tr(P_\alpha\cdot\{JP_\beta J\})              
\end{equation}
and we have a new quantum mechanics in a 3-dimensional octonionic Hilbert 
space. This is the exceptional Quantum Mechanical space discovered by Jordan, 
von Neumann and Wigner. In this case the octonions are real, so that the 
probabilities $\Pi_{\alpha\beta}$ are positive definite. If we take the 
special case $J=1$, then

\begin{equation}         
         \Pi_{\alpha\beta}=\Pi_{\alpha\beta}(1)={\rm Tr}(P_\alpha\cdot P_\beta)   
\end{equation}
corresponds to the square of the transition amplitude $<\alpha \mid \beta>$. If we 
put (see Mostow$^{\cite{most}}$)

\begin{equation}
{\rm cos}^{2}d_{\alpha\beta}={\rm Tr}(P_\alpha\cdot P_\beta)=\Pi_{\alpha\beta} ,
\label{eq:sekiz}      
\end{equation}
then, $d_{\alpha\beta}$ has a very simple geometric interpretation. The state 
$P_\alpha$ may be associated with a point with homogeneous coordinates 
$\alpha_1, \alpha_2, \alpha_3$ with the constraint Eq.(\ref{eq:sekiz}). The inhomogeneous 
coordinates

\begin{equation}
         x_1=\alpha_1\alpha^{-1}_3, \hspace{15pt} x_2=\alpha_2\alpha^{-1}_3  
\end{equation}
when $\alpha_3$ is self-conjugate, label a point in the two-dimensional 
octonionic projective plane as shown by Moufang$^{\cite{mou}}$ in 1933. Then it is 
easy to show that $d_{\alpha\beta}$ defined by Eq.(\ref{eq:sekiz}) is the 
non-euclidean distance between the points $P_\alpha$ and $P_\beta$ in the 
projective octonionic plane.
         
         Hence, in the octonionic Quantum Mechanics, a point represents a 
state associated with the projection operator $P_\alpha$ and the distance 
between two points is related to the transition probability between the 
corresponding states.
         
         The line in the projective plane defined by the points $P_\alpha$ 
and $P_\beta$ is associated with

\begin{equation}
         P_{\alpha\beta} = P_\alpha \times P_\beta                         
\end{equation}
         that obeys

\begin{equation}
         P^{2}_{\alpha\beta}=P_{\alpha\beta}                              
\end{equation}
         
         Lines and points obey all the postulates of projective geometry but 
do not satisfy the Desargues theorem. Hence the new Quantum Mechanical space 
corresponds to a new non-Desarguesian projective geometry.         
        
 The condition for a state $ \mid \gamma>$ to be the superposition of states 
$ \mid \alpha>$ and $ \mid \beta>$ in quantum mechanics can be reexpressed in 
geometrical language and in terms of the projection operators $P_\gamma$, 
$P_\alpha$  and $P_\beta$. Geometrically, the point $\gamma$ with homogeneous 
coordinates $ \mid \gamma>$ is on the line determined by the points $\alpha$ and 
$\beta$. It follows that the determinant of the matrix with columns $ \mid \alpha>$,
$ \mid \beta>$ and $ \mid \gamma>$ vanishes.  It is easily shown that this condition is equivalent to the relation

\begin{equation}         
   {\rm Tr}\{P_\gamma\cdot(P_\alpha \times P_\beta)\}=0     \label{eq:on}            
\end{equation} 
which uses only projection operators and the Jordan product$^{\cite{jp}}$. In this 
form the geometrical property can also be generalized to octonionic Quantum 
Mechanics, so that Eq.(\ref{eq:on}) expresses the condition for the state $\gamma$ to 
be a linear superposition of the states $\alpha$ and $\beta$.
         
         The Desargues property can now be rephrased in Quantum Mechanical
language as follows.
         
         Geometrically we have a point $S$ and 3 points $A$, $B$, $C$ not on a line. 
Take a point $A'$ on the line $SA$, a point $B'$ on the line $SB$ and a point $C'$ on 
the line $SC$ such that $A'$, $B'$, $C'$ do not lie on a line. Now the lines $AB$ and 
$A'B'$ intersect at a point $C"$. Similarly $BC$ and $B'C'$ intersect at $A"$ and $CA$ 
and $C'A'$ intersect at $B"$. The geometry is Desarguesian if $A"$, $B"$ and $C"$ lie 
on the same line. The Desargues property holds in a projective geometry if 
the projective plane has dimension $d>2$. In 2 dimensions it holds if the 
coordinates of the point are real, complex or quaternionic, i.e. if they 
belong to an associative algebra. On the other hand in the Moufang plane, 
the coordinates of a point are octonionic. Because of the non-associativity 
of octonions, there exist non-Desarguesian configurations.
         
         In Quantum Mechanics we take states corresponding to the points $S$, 
$A$, $B$, $C$. We superpose $S$ and $A$ to obtain $A'$ and proceed similarly for $B'$ and $C'$.
 We now construct a state $C"$ which is simultaneously a superposition of states 
$A$, $B$ and also or states $A'$, $B'$. We construct states $A"$ and $B"$ in the same way. 
If the states $A"$, $B"$, $C"$ are linearly related we have the Desargues property. 
If not, we say that we have a non-Desarguesian Quantum Mechanics. It follows 
that the octonionic Quantum Mechanics of Jordan, Wigner and von Neumann is 
non-Desarguesian. It represents a completely new kind of Quantum Mechanics 
in a finite dimensional Hilbert space in which the superposition principle 
is modified for states corresponding to non-associative triples.

Let us now look at the Desargues' theorem, making the following  octonionic entries in the places of the above letters. First of all, we see that from the entries of $\epsilon_{ijk}$'s that contain $e_1$ element are the $ijk$ combinations $123, 165$ and $174$. Accordingly, in the figure below we place $e_1$ in the place of $S$:
\vspace{3cm}
\begin{center}
\unitlength=.3mm
\begin{picture}(240,180)(0,0)
\put(102,10){\line(-1,3){53}}
\put(0,23){\line(1,3){49}}
\put(50,0){\line(0,3){170}}
%

%
\put(102,10){\circle*{3.7}}
\put(0,25){\circle*{3.7}}
\put(50,0){\circle*{3.7}}
\put(50,40){\circle*{3.7}}
\put(50,170){\circle*{3.7}}
\put(28,108){\circle*{3.7}}
\put(92,42){\circle*{3.7}}

%
\put(50,180){\makebox(0,0){$e_1$}}
\put(20,120){\makebox(0,0){$e_2$}}
\put(-10,20){\makebox(0,0){$e_3$}}
\put(60,45){\makebox(0,0){$e_6$}}
\put(40,-10){\makebox(0,0){$e_5$}}
\put(100,50){\makebox(0,0){$e_7$}}
\put(110,16){\makebox(0,0){$e_4$}}

%


\end{picture}
\end{center}
\vskip 0.5cm
\centerline{\bf Fig. 2~~~~~~~~~~~~~~~~~~~~~}

\vskip 1cm

Notice that $e_1 e_2=e_3$, $e_1 e_6= e_5$ and $e_1 e_7=-e_4$. 

Now we connect $e_2$ with $e_6$, and $e_3$ with $e_4$, and the point where the lines meet is marked $A$. Similarly, lines connecting $e_2$ with $e_7$, and $e_3$ with $e_4$ meet at $B$, and finally lines connecting $e_6$ with $e_7$ and $e_5$ with $e_4$ meet at $C$:
\vspace{1cm}
\begin{center}
\unitlength=.3mm
\begin{picture}(240,180)(0,0)
\put(102,10){\line(-1,3){53}}
\put(0,23){\line(1,3){49}}
\put(50,0){\line(0,3){170}}
\thicklines{
\put(67,-9){\line(-2,1){70}}
\put(66,-8){\line(-1,3){39}}
\put(50,0){\line(6,1){230}}
\put(50,40){\line(4,0){230}}
\put(-2,25){\line(6,-1){135}}
\put(28,108){\line(1,-1){105}} }

\put(101,10){\circle*{3.7}}
\put(0,25){\circle*{3.7}}
\put(50,0){\circle*{3.7}}
\put(50,40){\circle*{3.7}}
\put(50,170){\circle*{3.7}}
\put(28,108){\circle*{3.7}}
\put(93,42){\circle*{3.7}}
\put(66,-8){\circle*{3.7}}
\put(133,2){\circle*{3.7}}
\put(280,39){\circle*{3.7}}

\put(50,180){\makebox(0,0){$e_1$}}
\put(20,120){\makebox(0,0){$e_2$}}
\put(-10,20){\makebox(0,0){$e_3$}}
\put(60,45){\makebox(0,0){$e_6$}}
\put(40,-10){\makebox(0,0){$e_5$}}
\put(100,50){\makebox(0,0){$e_7$}}
\put(110,16){\makebox(0,0){$e_4$}}
\put(66,-18){\makebox(0,0){$A$}}
\put(135,-10){\makebox(0,0){$B$}}
\put(280,30){\makebox(0,0){$C$}}

%


\end{picture}
\end{center}
\vskip 1cm
\centerline{\bf Fig. 3~~~~~~~~~~~~~~~~~~~~~}
\vskip 1.5cm

We can now go ahead and replace point $A$ with $e_4$, since $e_3 e_5=e_4$ and $e_2 e_6=e_4$. Similarly since $e_2 e_7=e_3 e_4=e_5$ point $B$ is replaced by $e_5$, and finally $C$ is replaced by $e_3$ since $e_6 e_7=e_5 e_4=e_3$.
\vspace{2cm}
\begin{center}
\unitlength=.3mm
\begin{picture}(240,180)(0,0)
\put(102,10){\line(-1,3){53}}
\put(0,23){\line(1,3){49}}
\put(50,0){\line(0,3){170}}
\thicklines{
\put(67,-9){\line(-2,1){70}}
\put(66,-8){\line(-1,3){39}}
\put(50,0){\line(6,1){230}}
\put(50,40){\line(4,0){230}}
\put(-2,25){\line(6,-1){135}}
\put(28,108){\line(1,-1){105}} }

\put(101,10){\circle*{3.7}}
\put(0,25){\circle*{3.7}}
\put(50,0){\circle*{3.7}}
\put(50,40){\circle*{3.7}}
\put(50,170){\circle*{3.7}}
\put(28,108){\circle*{3.7}}
\put(93,42){\circle*{3.7}}
\put(66,-8){\circle*{3.7}}
\put(133,2){\circle*{3.7}}
\put(280,39){\circle*{3.7}}

\put(50,180){\makebox(0,0){$e_1$}}
\put(20,120){\makebox(0,0){$e_2$}}
\put(-10,20){\makebox(0,0){$e_3$}}
\put(60,45){\makebox(0,0){$e_6$}}
\put(40,-10){\makebox(0,0){$e_5$}}
\put(100,50){\makebox(0,0){$e_7$}}
\put(110,16){\makebox(0,0){$e_4$}}
\put(66,-18){\makebox(0,0){$e_4$}}
\put(135,-10){\makebox(0,0){$e_5$}}
\put(280,30){\makebox(0,0){$e_3$}}

\end{picture}
\end{center}
\vspace{1cm}
\centerline{ \bf Fig. 4~~~~~~~~~~~~~~~~~~~~~}
\vspace{1.5cm}
We notice that points $A, B, C$ (or, their replacements $e_4$, $e_5$ and $e_3$) are collinear (Desargues' theorem on a plane) and that all the points in the above figure are mapped onto a line $ABC=e_4 e_5 e_3$. Note that the lower triangle $\Delta~e_3 e_4 e_5$ is also mapped into line $ABC$. 

By replacing $e_1$ with any other $e_j$, $j\neq 1$, we can reproduce $7$ different combinations of triads associated with the full rotation through increments of an angle $2\pi/7$ of the dial described earlier.

We now proceed to Pappus' theorem, connect it with octonionic projective geometry described above and relate it to duality: We begin by drawing two arbitrary lines ($A$ and $B$) with arbitrarily picked points $a_1, a_2, a_3, a_4$ on $A$, and $b_1, b_2, b_3, b_4$ on $B$. We connect the first two pairs, $a_1$ to $b_2$, and $a_2$ to $b_1$, denoting the point where they cross by $P$. Similarly, connecting $a_1$ to $b_3$ and $a_3$ to $b_1$, we denote the point where they cross by $Q$, and connecting $a_3$ with $b_2$ and $a_2$ with $b_3$ we get the point $R$. Pappus' theorem tells us that these point $P$, $Q$ and $R$ are collinear. This is shown in the picture below: 
\vspace{3cm} 
\begin{center}
\unitlength=.3mm
\begin{picture}(240,180)(0,0)
%
\put(-80,100){\line(3,1){300}}
\put(-70,0){\line(3,-1){300}}
\put(-45,-10){\line(2,3){104}}
\put(20,-30){\line(-1,3){49}}
\put(20,-30){\line(3,5){123}}
\put(-45,-10){\line(1,1){185}}
\put(169,-79){\line(-1,1){197}}
\put(169,-79){\line(-1,2){112}} 

\put(-45,-10){\circle*{3.7}}
\put(20,-30){\circle*{3.7}}
\put(169,-79){\circle*{3.7}}
\put(-28,117){\circle*{3.7}}
\put(-6,48){\circle*{3.7}}
\put(28,62){\circle*{3.7}}
\put(58,146){\circle*{3.7}}
\put(88,82){\circle*{3.7}}
\put(140,173){\circle*{3.7}}
\put(170,183){\circle*{3.7}}
\put(215,-95){\circle*{3.7}}
\put(-28,130){\makebox(0,0){$a_1$}}
\put(58,159){\makebox(0,0){$a_2$}}
\put(140,186){\makebox(0,0){$a_3$}}
\put(170,196){\makebox(0,0){$a_4$}}
\put(-45,-23){\makebox(0,0){$b_1$}}
\put(20,-43){\makebox(0,0){$b_2$}}
\put(169,-92){\makebox(0,0){$b_3$}}
\put(215,-108){\makebox(0,0){$b_4$}}
\put(-90,100){\makebox(0,0){$A$}}
\put(-85,3){\makebox(0,0){$B$}}
\put(-16,48){\makebox(0,0){$P$}}
\put(28,48){\makebox(0,0){$Q$}}
\put(96,77){\makebox{$R$}}
%


\end{picture}
\end{center}
\vspace{5cm}
\centerline{ \bf Fig. 5~~~~~~~~~~~~~~~~~~~~}

\vspace{1.5cm} 

We can now replace the points $a_1=e_1$, $a_2=e_7$, $a_3=e_6$ and $a_4=e_2$ on the $A$ line, and $b_1=e_1$, $b_2=e_7$, $b_3=e_6$ and $b_4=e_2$ on the $B$ line. Notice that since $e_1 e_7=-e_4$, $P$ is replaced by $e_4$, and since $e_1 e_6=e_5$ we replace $Q$ by $e_5$, and because $e_7 e_6=-e_3$, $R$ is replaced by $e_3$. We notice because of Pappus' theorem, collinear points $e_4, e_5, e_3$ ($ \epsilon_{453} $) is dually mapped onto two arbitrary lines with entries $e_1, e_7, e_6, e_2$ (its dual $\psi_{6172}$). This is shown in the picture below:
\vspace{3cm}  

\begin{center}
\unitlength=.3mm
\begin{picture}(240,180)(0,0)
%
\put(-80,100){\line(3,1){300}}
\put(-70,0){\line(3,-1){300}}
\put(-45,-10){\line(2,3){104}}
\put(20,-30){\line(-1,3){49}}
\put(20,-30){\line(3,5){123}}
\put(-45,-10){\line(1,1){185}}
\put(169,-79){\line(-1,1){197}}
\put(169,-79){\line(-1,2){112}} 

\put(-45,-10){\circle*{3.7}}
\put(20,-30){\circle*{3.7}}
\put(169,-79){\circle*{3.7}}
\put(-28,117){\circle*{3.7}}
\put(-6,48){\circle*{3.7}}
\put(28,62){\circle*{3.7}}
\put(58,146){\circle*{3.7}}
\put(88,82){\circle*{3.7}}
\put(140,173){\circle*{3.7}}
\put(170,183){\circle*{3.7}}
\put(215,-95){\circle*{3.7}}
\put(-28,130){\makebox(0,0){$e_1$}}
\put(58,159){\makebox(0,0){$e_7$}}
\put(140,186){\makebox(0,0){$e_6$}}
\put(170,196){\makebox(0,0){$e_2$}}
\put(-45,-23){\makebox(0,0){$e_1$}}
\put(20,-43){\makebox(0,0){$e_7$}}
\put(169,-92){\makebox(0,0){$e_6$}}
\put(215,-108){\makebox(0,0){$e_2$}}
\put(-90,100){\makebox(0,0){$A$}}
\put(-85,3){\makebox(0,0){$B$}}
\put(-16,48){\makebox(0,0){$e_4$}}
\put(28,48){\makebox(0,0){$e_5$}}
\put(96,77){\makebox{$e_3$}}
%


\end{picture}
\end{center}
\vspace{5cm}
\centerline{ \bf Fig. 6~~~~~~~~~~~~~~~~~~~~~}
\vspace{1.5cm}
Therefore all other remaining $6$ triads of $\epsilon_{ijk}$'s are mapped onto their dual $\psi_{\alpha\beta\gamma\delta}$'s. 
         
         We now describe how two other features in quantum Mechanics can be 
rephrased in the Jordan formulation and hence generalized to octonionic 
Quantum Mechanics.
         
         The first concerns the insertion of a complete set of states$^{\cite{jp}}$.  If $I$ denotes the unit matrix, we have

\begin{equation}         
         I=\sum_{i}^{} \mid i><i \mid =\sum_{i}^{}P_i,                            
\end{equation}
         so that the relation

\begin{equation}         
         <\alpha \mid \beta>=\sum_{i}^{}<\alpha \mid i><i \mid \beta>             
\end{equation}
can be transcribed as

\begin{equation}        
         \Pi_{\alpha\beta}={\rm Tr}(P_\alpha\cdot P_\beta)=\sum_{i}^{}\{P_\alpha 
P_i P_\beta\}     
\end{equation}
which is also valid in octonionic quantum mechanics.
         
         The second is related to the Jordan formulation of compatible 
observables. In the associative case we have an identity that relates the 
associator to the double commutator, namely

\begin{equation}      
         [AJB]=-\frac{1}{4}[[A,B],J],         \label{eq:onbir}       
\end{equation}
so, that if $J$ is arbitrary and $A$, $B$, $C$ are hermitian the equation

\begin{equation}         
         [AJB]=0           \label{eq:oniki}                                 
\end{equation}
implies

\begin{equation}         
         [A,B]=0.                              
\end{equation}
  Since Eq.(\ref{eq:onbir}) does not hold in the octonionic case, we define the 
compatibility of the observables $A$ and $B$ by Eq.(\ref{eq:oniki}) for arbitrary 
$J$. This is a condition which only involves the Jordan product of octonionic 
observables.

\section*{Invariance Properties of the Exceptional Ouantum Mechanics ($F_4$). Generalization to Complex Jordan Algebras ($E_6$).}
         
         In the case of the usual Quantum Mechanics transition amplitudes are 
invariant under unitary transformations as

\begin{equation}         
         <\alpha' \mid \beta'>=<\alpha \mid \beta>                     
\end{equation}
when

\begin{equation}
         \mid \alpha'>=U \mid \alpha>, \hspace{15pt} \mid \beta'>=U \mid \beta>, \hspace{15pt} 
U U^{\dagger} =1
\end{equation}
Then the projection operators transform as

\begin{equation}         
               P'_\alpha =UP_\alpha U^{\dagger},~~~~~~   P'_\beta=U P_\beta 
U^{\dagger}.                       
\end{equation}
The observables $\Omega$ that are linear combinations of projection operators 
also transform in the same way

\begin{equation}
         \Omega'=U\Omega U^{\dagger}          
\end{equation}
so that the Jordan product $\Omega$ of two observables $\Omega_1$ and 
$\Omega_2$ also transforms like a projection operator, since

\begin{eqnarray}      
\Omega'&=&\Omega'_1\cdot\Omega'_2=\frac{1}{2} (\Omega'_1\Omega'_2 + \Omega'_2\Omega'_1) \nonumber \\
&=& \frac{1}{2}U(\Omega_1\Omega_2+\Omega_2\Omega_1)U^{\dagger}=U \Omega 
U^{\dagger}
\end{eqnarray}
         
         It follows that, in a $n$-dimensional Hilbert space, with $n \times n$ 
hermitian matrices associated with observables and projection operators 
for states, the automorphism group of the Jordan algebra of observables 
is $U(n)$ or $SU(n)$. In order to find the invariance group of octonionic 
Quantum Mechanics we must find the automorphism group of the exceptional 
Jordan algebra. This was shown to be the group $F_4$ by Chevalley and 
Schafer$^{\cite{cs}}$ more than a decade after the discovery of exceptional 
Jordan algebras.
         
         The infinitesimal action of $F_4$ on an element $J$ of the Jordan 
algebra can be written simply.
         
         If $H_1$ and $H_2$ are traceless octonionic hermitian matrices we
have$^{\cite{grs}}$

\begin{equation}
         \delta J=[H_1, J, H_2]           
\end{equation}
         The transformation property of the projection operators $P_\alpha$ 
for  states $\alpha$ is obtained by putting $J = P_\alpha$.  Let us show that 
this gives the unitary group in the associative case. Let

\begin{equation}
         iH=-\frac{1}{4}[H_1, H_2], \hspace{15pt} H=H^{\dagger}               
\end{equation}
	Then, using Eq.(\ref{eq:onbir}) we can write

\begin{equation}
         \delta J =i[H, J]                                                    
\end{equation}
Exponentiation gives

\begin{equation}
J'=J+[iH,J]+\frac{1}{2!}[iH,[iH,J]]+\ldots = e^{iH}~ J~ e^{-iH}              \label{eq:onuc}
\end{equation}
which shows that $J$ is transformed by a unitary matrix. In the octonionic 
case the finite transformation of $F_4$ is given by the series

\begin{equation}
         J'=J+[H_1,J,H_2]+\frac{1}{2!}[H_1,[H_1,J,H_2],H_2]+....     
\label{eq:ondort}
\end{equation}
which only involves the Jordan product and can no longer be written in the 
form Eq.(\ref{eq:onuc}). When $H_1$ involves only one octonion and $H_2$ is a purely scalar 
matrix then Eq.(\ref{eq:ondort}) can be integrated in the form Eq.(\ref{eq:onuc}) 
with $iH$ replaced by an 
antihermitian octonionic matrix involving one octonion only. It is seen that 
the full group is determined by the traceless hermitian matrices $H_1$ and 
$H_2$ so that it has 52 parameters. The invariants under the $F_4$ 
transformation are

\begin{equation}
         I_1={\rm Tr}J,       \label{eq:yirmi}                  
\end{equation}

\begin{equation}
         I_2={\rm Tr} J^{2},                      
\end{equation}

\begin{equation}
         I_3 = {\rm Det} J = \frac{1}{3} {\rm Tr} (J \cdot  J \times J )    
\end{equation}
         An irreducible representation of $F_4$ is obtained by taking $I_1$=0. 
It corresponds to traceless Jordan matrices.  We have seen that octonionic 
Quantum Mechanics based on real octonions provides us automatically with a 
finite Hilbert space with $F_4$ symmetry. Since $F_4$ has 
$SU(3) \times SU(3)^{c}$ 
as a maximal subgroup we have a fundamental justification for the color degree 
of freedom and the SU(3) flavor$^{\cite{h}}$. Under this group we have the 
decomposition 

\begin{equation}
26=(8,1) +(3,3)+(\bar{3}, \bar{3})   \label{eq:ykbes}
\end{equation}
The color singlet part which is a SU(3) flavor octet 
lies in an ordinary quantum mechanical space with SU(3) symmetry involving 
one of the octonionic imaginary units while the colored degrees of freedom 
involve the remaining six imaginary units.
         
The behavior of various states under the color group is best seen if 
we use split octonion units defined in Eqs.(\ref{eq:ellisekiz},\ref{eq:ellidokuz}).

We can now consider the general element $F$ of the Jordan algebra 
with complex components. It can be decomposed as follows

\begin{equation}
         F=u_0 L+u^{*}_0 L^{T} + u^{*}_j Q_j +u_j R^{*}_j  \label{eq:onyedi}   
\end{equation}
Here $L$, $Q_j$, $R_j$ are $3 \times 3$ complex matrices, $T$ denotes transposition and  
$Q_j$ and $R_j$ are antisymmetric so that

\begin{equation}
         Q_j=-Q^{T}_j, \hspace{15pt} R_j=-R^{T}_j    
\end{equation}
         
         If we associate $L$ with the ($\bar{3}$,3) representation of a group 
$SU(3) \times SU(3)$, $Q_j$ with (3,1) and $R^{*}_j$ with (1,$\bar{3}$), then, 
together with the color index $j$ we find that $F$ has the 
$SU(3) \times SU(3) \times SU(3)^{c}$ decomposition

\begin{equation}
         F=(\bar{3},3,1^{c})+(3,1,3^{c})+(1,\bar{3},\bar{3}^{c})    
\end{equation}

Comparison with Eq.(\ref{eq:ykbes}) tells us that $F$ is the 27-dimensional 
representation of the exceptional group $E_6$. The color singlet part $L$ 
can be associated with the lepton matrix $L$, that is, in terms of the $SU(3)\times SU(3)$ flavor group the leptons fall in a $(3\times 3)$ matrix that represents the $(\bar{3},3)$:

\begin{equation}
(\bar{3},3): \quad L^{(e)}= \left( \begin{array}{ccc}
\hat{N}_R  & \hat{\tau}_R & \hat{e}_R  \\
\hat{\tau}_L & \hat{\nu}_L & \beta_L^e \\
e_L & \nu_L^e & \alpha_L^e
\end{array} \right)  \label{3.2.8}
\end{equation}
Meanwhile, $Q_j$ and 
$R^{*}_j$ respectively are associated with left-handed quarks and right-handed antiquarks: the (3,1) quarks and the $(1, \bar{3})$ antiquarks are

\begin{equation}
(3,1): \quad \left( \begin{array}{c}
u_L^i \\ d_L^i \\ b_L^i
\end{array} \right), ~~~~\quad (1,\bar{3}):\quad
\left( \begin{array}{c}
\hat{u}_R^i \\ \hat{d}_R^i \\ \hat{b}_R^i
\end{array} \right)
\end{equation}
It follows that lepton and colored quark fields can be 
combined in a complex Jordan matrix of the form Eq.(\ref{eq:onyedi}) which is hermitian with respect to octonionic conjugation only, so that

\begin{equation}
         \bar{F}^{T}=u^{*}_0 L^{T} + u_{0} L -u^{*}_j Q^{T}_j -u_j 
{R^{*}_j}^{T}=F        
\end{equation}
         
         The $\bar{27}$ representation of $E_6$ corresponds to $F^{*}$.
         
         The $E_6$ transformation of $F$ involves three traceless real 
octonionic Jordan matrices $H_1$ ,$H_2$, $H_3$ and we have$^{\cite{h}}$

\begin{equation}
         \delta F=[H_1,F,H_2]+iH_3\cdot F          
\end{equation}
         
         This shows that $F_4$ is a subgroup of $E_6$ and that $E_6$ has 
$3 \times 26$=$78$ real parameters. The Freudenthal product of $F_1$ and $F_2$ 
projects out the $\bar{27}$ representation out of the symmetric product 
of $27 \times 27$, so that we can write

\begin{equation}
   F_1 \times F_2 = F^{*}_3, \hspace{15pt}  F^{*}_1 \times F^{*}_2 =F_3       
\end{equation}
         
         Another $E_6$ invariant operation is the triple product defined 
by Eq.(\ref{eq:yirmi}), so that

\begin{equation}
 \{F^{*}_1 F_2 F^{*}_3\}=F^{*}_4,\hspace{15pt} \{F_1 F^{*}_2 F_3\}=F_4    
\end{equation}
are also invariant if $F_1, F_2, F_3, F_4$ transform like (27). Finally we 
can construct the invariant

\begin{equation}
         (F_1, F_2) = {\rm Tr} (F_1.F^{*}_2)                
\end{equation}
It follows that, with three Jordan matrices $F_1$, $F_2$, $F_3$ we can associate 
the invariant

\begin{equation}
         (F_3, F_1 \times F_2)={\rm Tr}(F_3 \cdot F_1 \times F_2)         
\end{equation}
         
	Given one $F$, ${\rm Tr}~F$ is not $E_6$ invariant. But we can construct 4 
invariant real quantities $I_2, I_3$ and $I_4$ defined by

\begin{equation}         
 I_2=(F,F^{*}),\hspace{15pt} I_2+iI_3=(F,F\times F)=3{\rm Det}F
\end{equation}

\begin{equation}        
         I_4=(F\times F,F^{*} \times F^{*})          
\end{equation}

         A geometry which generalizes the projective geometry of the 
Moufang plane can be based on the complex matrices $F$. It is called the 
geometry of complex octonionic planes$^{\cite{fau}}$.  A generalized point (or 
state) is defined by $S$ such that

\begin{equation}         
         S \times S = 0                    
\end{equation}
         
         The distance $d_{12}$ between points $S_1$ and $S_2$ or the 
transition probability $\Pi_{12}$ is given by

\begin{equation}        
        \Pi_{12}={\rm cos}^{2}d_{12}=(S_1,S_2)={\rm Tr}(S_1\cdot S^{*}_2) 
\end{equation}
and is $E_6$ invariant.
         
         It is possible to associate idempotent projection operators with 
such states and generalize the quantum mechanical formalism. The geometry 
is more complicated than the Moufang geometry and all its quantum mechanical 
implications have not yet been worked out. However, the existence of this 
$E_6$ invariant exotic geometry and its close correspondence with the 
phenomenological symmetries of quarks and leptons as reviewed in the 
preceding section provides a strong motivation for the reformulation of the 
properties of the complex octonionic planes in purely quantum mechanical terms.
         
         If it turns out that $F_4$ or $E_6$ describe correctly the internal 
symmetries of fundamental fields we may seek the origin of these symmetries 
in the properties of unique finite Hilbert spaces associated with exotic 
geometries.

\section*{Conclusions}

The symmetries of the hadronic spectrum and the hadronic decays have 
uncovered a colored quark substructure. Weak and electromagnetic interactions 
showed us that quarks behave like leptons and a local field theory of both 
leptons and quarks makes sense. Strong interactions are well described by a 
local gauge theory based on the exact color group while weak and 
electromagnetic interactions are unified within a gauge field theory of 
spontaneously broken local flavor group. The symmetries between leptons 
and quarks led us to the notion that these fundamental fermions belong to 
the same multiplet of a unifying group. The successful candidates for such 
a unifying group have turned out to be subgroups of the exceptional group 
$E_6$. On the other hand, the only non-trivial generalizations of the Hilbert 
space of Quantum Mechanics and the algebra of observables involve algebraic 
and geometrical structures connected with the exceptional groups $F_4$ and 
$E_6$. These unique and intrinsically finite structures exhibit an exact 
color symmetry originating in octonions that go in the building of these 
exotic structures.
         
	 The fundamental symmetries of elementary 
particles seem to point to field theories based on local internal symmetries 
connected with structures that can be constructed by using octonions.
         
         The finite octonionic structures are grafted to each point of 
space-time, or to each state of an infinite Hilbert space in which the 
Poincar\'e group or the conformal group act unitarily. The octonionic structure 
at a point $x$ is related to the one at point $x = 0$ by an automorphism of 
this algebraic structure. Hence the parameters of the automorphism must be 
functions of $x$ and we must have local invariance under the automorphisms of 
the octonionic structure$^{\cite{jp}}$. This introduces a connection which 
describes how the automorphism at point $x + dx$ is obtained from the one at $x$. As a result the 
internal space and the $x$ space become connected in a fibre bundle structure 
with the Hilbert space of the external group as a base and the observables of 
the charge space or the generators of the automorphism of their algebra as a 
fibre. This provides the geometrical picture associated with a local gauge 
field theory based on the automorphism algebra of the internal octonionic 
observables.
         
Octonionic observables form a Jordan algebra. Then the automorphism group is an 
exceptional group $F_4$ or $E_6$, and a gauge field theory of 
quarks and leptons based on exceptional groups emerges as shown by G\"ursey and collaborators. It is also possible to 
consider super Jordan algebras for generalized Jordan algebras involving 
both bosonic and fermionic observables$^{\cite{fglm}}$. In this case the automorphism 
group is a supergroup. If the generalized Jordan algebra is octonionic then 
its automorphism is given by an exceptional supergroup. In the case the 
supergroup is simple it can be the supergroup $G(3)$ or the supergroup 
$F(4)$ $^{\cite{kac}}$. The latter has 40 parameters and its Lie subgroup is 
$SO(7) \times SU(2)$ which admits the phenomenological 
$SU(2) \times U(1) \times SU(3)^{c}$ 
as subgroup. Here again SO(7) is the normal group of purely imaginary 
octonions and SU(3) is the subgroup of the $G_2$ automorphism of the 
octonionic algebra. If the $E_6$ model becomes untenable phenomenologically 
the super $F(4)$ is another possible candidate. Instead of leptoquarks and 
diquarks it would introduce spin 3/2 gauge particles alongside with gauge 
bosons, so it would be similar to supergravity when used as a local 
supersymmetry group.
         
         Many problems still remain to be solved. The outstanding ones are concerned with the building of Fock space (tensor 
products with exceptional structures), the understanding of color confinement 
and the realization of the ultimate synthesis with supergravity. Then we might
also be able to understand the emergence of a superheavy mass scale that 
seems necessary for the unification of strong interactions with 
electromagnetic and weak interactions and for the explanation of large 
renormalization shifts in experimentally accessible parameters like the 
Weinberg angle as well as for the quasi-stability of the proton.


\newpage

\begin{thebibliography}{99}

\bibitem{gurtze} C-H. Tze and F. G\"ursey, On the Role of Division, Jordan and Related Algebras in Particle Physics, World Scientific, 1996.
\bibitem{yokota} I. Yokota, J. Fac. Sci. Shinsu Univ. {\bf 3} (1968) 61.
\bibitem{catmor} S.Catto and C.J. Moreno, to be published.

\bibitem{1} S. Catto and F. G\"ursey, Nuovo Cimento {\bf A86} (1985) 201.

\bibitem{2} S. Catto and F. G\"ursey, Nuovo Cimento {\bf A99} (1988) 685.

\bibitem{3} M. Anselmino, et al. Rev. Mod. Phys. {\bf 65} (1993) 1199.
\bibitem{4} E. Klempt, "Baryon Resonances and Strong QCD," arXiv:nucl-ex/0203002.
 

\bibitem{6} F. Iachello, Proceedings of ICGTMP, Ed. XX, 1992(?), Ciemat Publ.

\bibitem{7} J. Baez, Bull. Am. Math. Soc. {\bf 39(2)} (2002) 145-205.

\bibitem{fgmg} M. G\"unayd{\i}n and F. G\"ursey, J. Math. Phys. {\bf 14} (1973) 1651.



\bibitem{sc} S. Catto, Symmetries in Science, {\bf VI} (1993) 129. Ed. B. Gruber, Plenum Press. 

\bibitem{jor} P. Jordan, {\em{Nachr. Ges. Wiss. G\"{o}ttingen, (1933) 209.}}

\bibitem{jp} F. G\"{u}rsey, "Non-associative Algebras in Quantum Mechanics and
Particle Physics." U. Virginia, 1977.

\bibitem{jnw} P. Jordan, J. von Neumann and E.P. Wigner, Ann. Math. {\bf 35} (1934) 29.

\bibitem{alb}A.A. Albert, Ann. Math. {\bf 35} (1934) 65.

\bibitem{freud}H. Freudenthal, {\em{Adv. in Math. I, 145 (1965)}}.

\bibitem{most}G.D. Mostow, "Strong Rigidity of Locally Symmetric Spaces."
(Princeton U. Press, 1973).

\bibitem{mou}R. Moufang, Abh. Math. Sem. Univ. Hamburg {\bf 9} (1933) 207.

\bibitem{cs}C. Chevalley and R.D. Schafer, "An Introduction to Non-Associative 
Algebras." (Academic Press, New York 1966).

\bibitem{grs}M. G\"{u}nayd{\i}n and F. G\"{u}rsey, Phys. Rev. {\bf D9} (1974) 3387.

\bibitem{h}F. G\"{u}rsey, P. Ramond and P. Sikivie, Phys. Lett. {\bf 60B} (1976) 177.

\bibitem{fau}J.R. Faulkner, {\em Memoirs of the Am. Math. Soc. No.104 (Providence,
RI 1970)}.

\bibitem{fglm}F. G\"{u}rsey and L. Marchildon, J. Math. Phys. {\bf 19} (1977) 942.
 
\bibitem{kac}V. Kac, Comm. Math. Phys. {\bf 53} (1977) 31.
\end{thebibliography}
\end{document}